\documentclass[slac_one]{revtex4}
\usepackage{graphicx}
\usepackage{fancyhdr}
\begin{document}

\title{Commissioning and Performance of the ATLAS Liquid Argon
  Calorimeters} %% Paper title goes here


\author{Dominik Dannheim\footnote{email: dominik.dannheim@cern.ch} 
on behalf of the ATLAS Liquid Argon Calorimeter Group}
\affiliation{CERN, CH-1211 Geneva, Switzerland}
\begin{abstract}
The ATLAS liquid argon (LAr) calorimeter system consists of an
electromagnetic barrel calorimeter and two end-caps with
electromagnetic, hadronic and forward calorimeters. 
The construction of the full calorimeter system is
completed since mid-2004.  The detector has been operated with LAr
at nominal high voltage and fully equipped with readout electronics. 
Online software, monitoring tools and offline signal reconstruction
have been developed for data collection and processing.
Extensive tests with calibration pulses have been carried out, and the 
electronics calibration scheme for all 182468 channels has been exercised.
Since Augst 2006, cosmic muon data have been collected together with 
the rest of the ATLAS detector system as part of the ATLAS
commissioning program. The reconstructed 
LAr signals from energy deposited by cosmic rays are compared to the 
prediction derived from measured detector parameters and calibration
pulses.   
The uniformity of the detector response within regions that have sufficient
cosmic muons are examined.
The expected performance of the LAr calorimeter for ATLAS physics, based on 
previous beam tests and Monte Carlo simulation, is also summarised. 
\end{abstract}

\maketitle

\thispagestyle{fancy}

\section{THE ATLAS LIQUID ARGON CALORIMETERS}
ATLAS~\cite{atlas-tdr,atlas-det} is a general-purpose detector 
built for operation at the Large Hadron Collider (LHC) at CERN. The 
LHC is a proton-proton collider which will operate with a 
center-of-mass energy of 14TeV. 
A system of LAr calorimeters located in three separate cryostats
between the inner tracking detectors and the outer muon chambers 
forms one of the major ATLAS detector systems~\cite{atlas-lar-tdr}. 

\subsection{Physics Requirements}
The requirements for EM calorimetry in ATLAS are driven by the discovery
potential of the LHC~\cite{atlas-calo-perf-tdr}. Major goals in the
design of the LAr calorimeters have been sensitivity to the
electromagnetic Higgs decay channels H$\rightarrow \gamma \gamma$ and
H$\rightarrow$ee as well as the ability to detect an
electromagnetically decaying exotic heavy resonance like a 
hypothetical new neutral gauge boson (Z$'$) or an excited
graviton ($G^*$). Furthermore, the EM calorimeters are required to
ensure sensitivity to a wide range of supersymmetric scenarios.

Full coverage in azimuthal angle $\phi$ and EM coverage up to
pseudorapidities of $|\eta|<3.2$ ($|\eta|<2.5$ with high precision)
without insensitive regions is achieved through an accordion geometry.
Good energy resolution with a stochastic term of
$<10\%/\sqrt{\mathrm{GeV}}$,
a noise term of $<300\mathrm{MeV}$ and a constant term of $<0.7\%$
requires precise mechanics and electronics calibration. Precision
measurements like the determination of the mass of the $W$ boson ask
for a linearity of better than $0.1\%$, which requires a very good
understanding and correction of dead material effects. Particle
identification (separation of electrons from jets and photons from
single $\pi_0$) is crucial for many physics signatures and requires
a lateral and longitudinal segmentation and a fine granularity in the first 
sampling layer. A rejection of single $\pi_0$ of better than 3 for
transverse momenta above 50~GeV is achieved with an angular resolution
of better than 50~mrad/$\sqrt{\mathrm{E}}$. The ability to detect
electromagnetically decaying heavy resonances at the kinematic limit
sets the dynamic range from about 30~MeV up to 3~TeV, corresponding to an
effective readout resolution of about 17~bits. To minimise pile-up
effects from overlay events at the LHC collision rate of 40~MHz
and at the LHC design luminosity of 
$10^{34}$cm$^{-2}$s$^{-1}$, a fast response of the readout is
necessary, which is achieved by using small readout gaps and bipolar
shaping. The LAr calorimeters are also expected to contribute
to the measurement of hadronic and missing energy in the forward
region. For this purpose, hadronic and forward calorimeter systems
have been implemented, extending the coverage up to $|\eta|<4.9$ with
an energy resolution of $\sigma _E /E\approx 50\%/\sqrt{\mathrm{E}} \oplus 3\%$
for $|\eta| <3$ and $\sigma _E /E\approx 100\%/\sqrt{\mathrm{E}} \oplus 10\%$
for $|\eta| >3$.
\subsection{Electromagnetic Calorimeters}
A Pb/LAr sampling technique with an accordion geometry was
chosen both for the electromagnetic barrel calorimeter (EMB) and 
for the electromagnetic endcap calorimeter (EMEC). Using LAr as the active medium
results in inherent linearity, radiation hardness and a fast buildup of
the ionisation signal. Cu edging of the readout electrodes allows for
a small gap width of $\approx 1-3$~mm, leading to a drift time of approximately 450~ns.
A nominal high voltage of 1~kV/mm is applied, with a two-fold
redundancy per readout electrode.
A total of 173312 readout channels is distributed over 4 longitudinal
layers covering the region up to $|\eta|<1.475$ (EMB) and $1.375 < |\eta| <
3.2$ (EMEC). 
A presampler (for $|\eta|<1.8$) measures the energy loss
in the material in front of the active calorimeters and inside the
cryostats, which corresponds to
approximately 2-6 radiation lengths $X_0$. It is followed by the
finely segmented strips layer used for discrimination between photons and single $\pi_0$
(granularity from $\Delta \eta \times \Delta \phi=0.003 \times 0.1$ to
$0.1 \times 0.1$ and depth of $\approx 4 X_0$). The middle layer contains the bulk of the EM shower
energy ($17-20 X_0$). The back layer contains the remaining part of the EM shower
($2-10 X_0$) and is used for rejection of hadronic showers.

\subsection{Hadronic Endcap Calorimeter}
A Cu-Ar sampling technique is used for the hadronic endcap
calorimeter (HEC), which consists of two
wheels (HEC1 and HEC2) in each endcap cryostat and covers the range
$1.5<|\eta|<3.2$. Double gaps on either side of the Cu readout electrode
with a total width of approximately 8.5~mm lead to a sampling fraction of
4.4\% (HEC1) and 2.2\% (HEC2). Low noise GaAs preamplifiers mounted inside the
cryostats are used to readout the signals of the 5632 channels. The
granularity in $\Delta \eta \times \Delta \phi$ is $0.1 \times 0.1 /
64$ (for $|\eta|<2.5$) and $0.1 \times 0.2$ (for $2.5<|\eta|<3.2$).  

\subsection{Forward Calorimeter}
The forward calorimeter (FCal) is composed of three modules featuring
cylindrical electrodes (rods) with thin liquid argon gaps. An electromagnetic
module with Cu absorber is followed by two hadronic modules with
tungsten absorbers, covering the range from $3.1<|\eta|<4.9$ with
3524 readout channels. The charges are collected on thin tubes maintained
between the absorber matrix and the rods, which allow to reduce the
gap  widths to approximately 0.25~mm to 0.5~mm and therefore reduce
the effect of ion buildup in this high-flux region.

\subsection{Detector Readout and Calibration}
Both the frontend-readout and the electronics calibration systems are mounted
on detector and must therefore be able to tolerate significant
levels of radiation. 
The on-detector location also implies that access
to the frontend electronics is limited to the long shutdown periods
foreseen in yearly intervals and therefore reliability of the readout
and calibration system is a key concern.
This holds in particular for the first stage of amplification of the
HEC readout signals, which is performed by GaAs preamplifiers mounted
in an inaccessible position inside the cryostats.

Frontend boards are used for preamplification, bipolar
shaping in 3 separate gain scales and analog storage of the detector
signals during the level-1 trigger latency of $\approx 2.5~\mu$s~\cite{feb-nim}. The
analog signals are digitised at the LHC bunch-crossing frequency of
40~MHz on the frontend boards and sent off-detector to the readout drivers
through optical links. During LHC running 5 time samples of 25~ns
are readout for each event. For commissioning studies up to 32 samples can be readout.
The frontend boards also provide the first
stage of summation of the analog signals used for the level-1 trigger.

The readout drivers calculate energy, time and a $\chi ^2$-like
quantity characterising the quality of the waveform~\cite{backend}. 
Furthermore, they provide
energy-weighted position moments for the missing transverse energy
calculation performed by the higher-lever trigger as well as
monitoring histograms. 

An electronics calibration system is used to characterise and monitor
the response of the detector cells~\cite{calib}. Charges are injected very close to
the electrodes, resulting
in a signal shape close to the ionisation signal. For the FCal a
simpler calibration scheme was chosen, where the calibration signals
are injected directly into the frontend-readout boards.
The dynamic range of
the calibration system is 16~bits and the observed non-linearity is
less than 0.1\%. Optimal filtering coefficients are obtained from the
calibration data taken in between LHC fills, 
which are used during physics running to calculate 
the energy online in the readout drivers.

\subsection{Commissioning with Calibration Measurements}
The construction and installation of the liquid argon calorimeter and its readout
system has been completed in March 2006. The calibration and readout
system have been used extensively during all stages of installation
and commissioning, to support the debugging and repair of hardware
problems. Monitoring of the readout status continues after 
loss of access to the frontend electronic with the closure of the
detector in June 2008. The current situation is satisfactory:

 \emph{Dead readout channels:} Only 0.02\% of all channels inside
  the detector are found to show no readout signal. No
  continuous dead regions are observed. For about 0.9\% of all channels 
  the frontend readout is currently not functioning. These recent problems are
  expected to be fixed in the shutdown following the first beam
  commissioning period of the LHC machine.

 \emph{Problematic channels:} About 0.5\% of all channels show
  minor problems like increased noise or damaged calibration lines.
  Offline corrections are applied in these cases and the impact on the
  calorimeter performance is negligible. No continuous problematic regions are observed.

 \emph{High-voltage status:} Less than 1\% of all HV channels
  are operated at reduced voltage. The voltage is sufficient to
  guarantee usable signals for all readout electrodes. 

\section{COSMIC DATA TAKING}
Cosmic muon signals are the first and only physics data before
LHC operation starts. The LAr calorimeters take cosmic runs since 2006,
with a few $10^7$ events analysed so far. Dedicated cosmics triggers based on
the hadronic tile calorimeter ($\Delta \eta \times \Delta \phi = 0.1
\times 0.1$) and more recently other triggers
(e.g. the level-1 calorimeter trigger) have been used. 
Two different selections are applied for the offline analysis of the
data:
1) Samples of high-energy muons with reconstructed cluster energies above 500~MeV
are used to check the quality of the physics-pulse prediction. The
limited acceptance and statistics of this selection ($\approx$1\% of
the total cosmics yield) allow only for qualitative studies.
2) Minimum ionising particles (MIPs) are selected to check the EM
performance in terms of uniformity and timing. The observed signals
are mostly small and non-projective.

\subsection{Quality of Physics-Pulse Prediction}
\begin{figure*}[t]
\centering
\includegraphics[width=8cm]{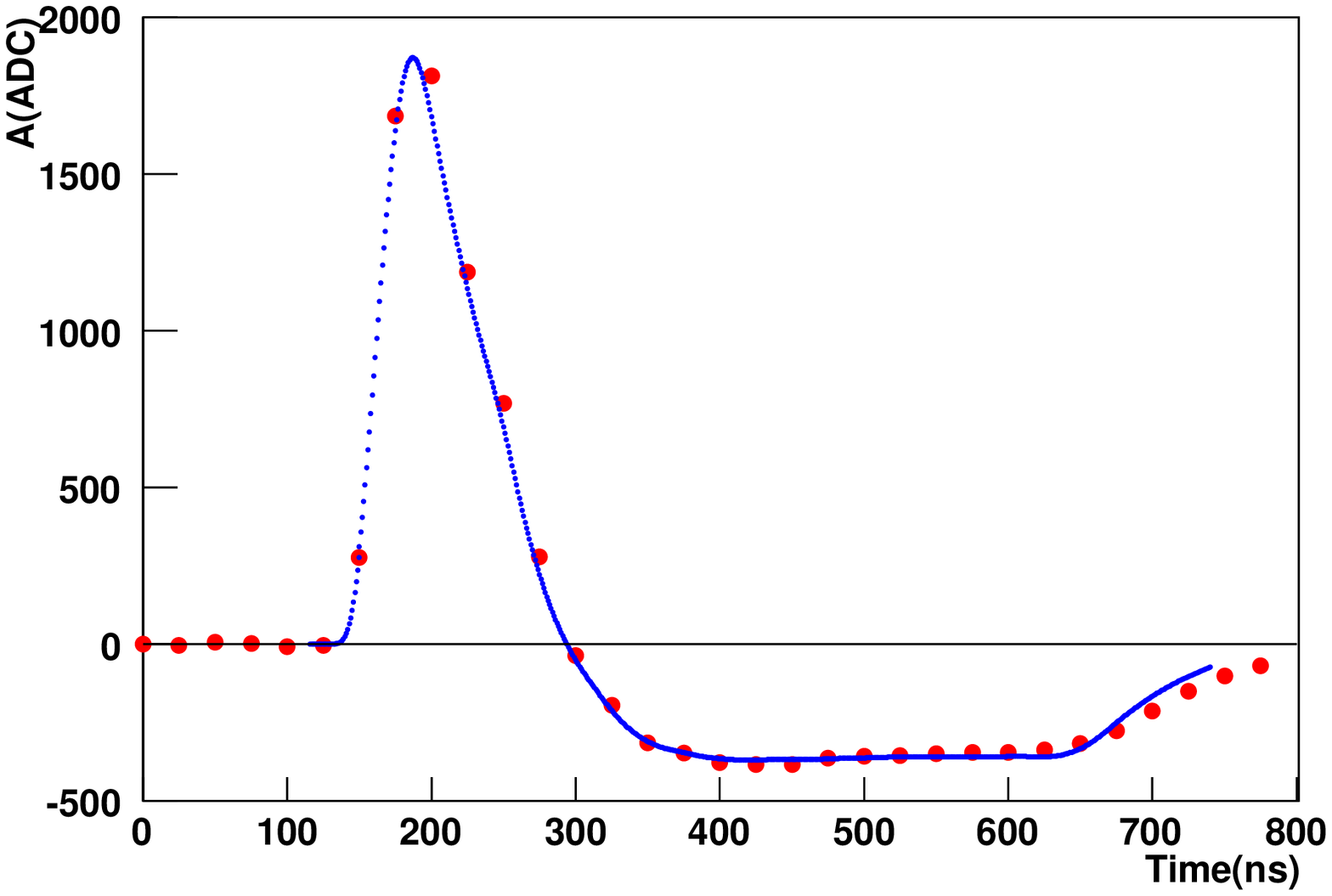}
\hspace{0.5cm}
\includegraphics[width=8cm]{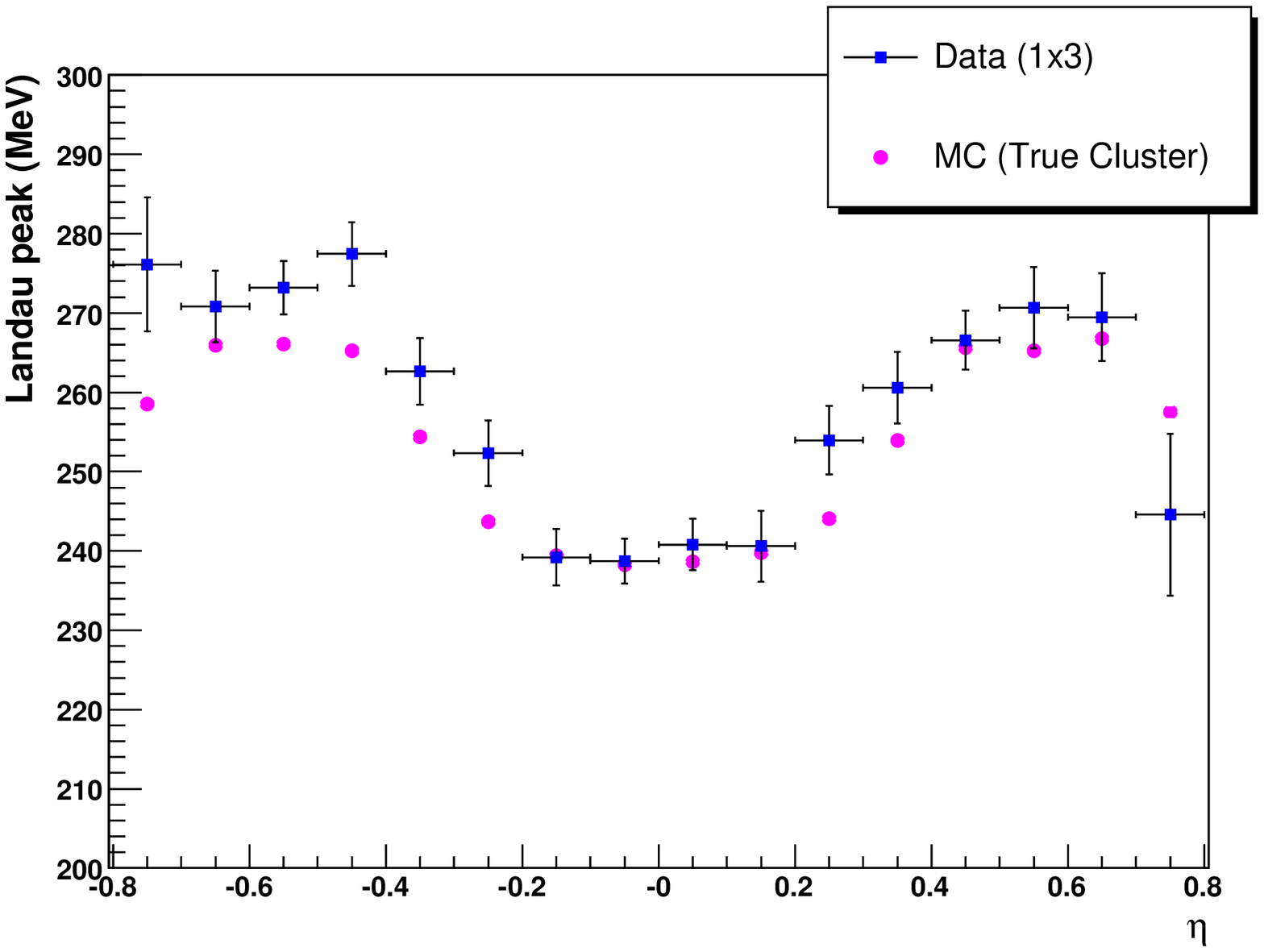}
\begin{picture}(0,0)
\put(-480, 160){a)}
\put(-220, 160){b)}
\end{picture}
\caption{a) Observed physics pulse (dots) and prediction from
  factorisation of the calibration readout response (solid line).
b) Response non-uniformity for a selection of cosmic muon
  events as function of pseudorapidity. The data (dots with error
  bars) are compared to the prediction from a MC simulation (dots).} \label{fig-cosmic}
\end{figure*}

High-energy muons (E$>500$~MeV) are selected and the observed physics
pulses are compared to the prediction from a factorisation of the
calibration-readout response (Fig.~\ref{fig-cosmic}a). Good agreement 
and a coherent quality of
the signal-reconstruction is observed for the 5 highest samples 
over the whole range of the EM calorimeters
($|\eta|<3.2$). Small deviations of the observed signals from the
prediction for the later samples of the pulse are understood to be
caused by variations in drift time due to displacements of the 
readout electrodes between the absorbers from their nominal positions.
The good understanding of the readout response leads to very small
systematic effects from the signal reconstruction. 

\subsection{Uniformity Studies with MIPs}
A more inclusive selection of projective MIPs in the region
$|\eta|<0.8$ was used to study the
uniformity of the calorimeter response~\cite{cosmic-note}. A 1x3 cells clustering
algorithm was used for the reconstruction of the calorimeter energy
and compared to a Monte Carlo (MC) simulation. A landau distribution
of the reconstructed energy was observed and the agreement between
data and MC for the response non-uniformity was found to be 
within 3\% over the whole $\eta$ range (Fig.~\ref{fig-cosmic}b).

\section{PERFORMANCE IN TESTBEAM}
The expected performance of the EM calorimeter is estimated from
results of a combined testbeam in 2004. A full slice of the barrel
detector was placed in a magnetic field of 1.4~T and subject to
particle beams of electrons, photons, pions, protons and muons from 1
to 350~GeV. The results have been used to validate the calorimeter
simulation and to extract performance parameters like resolution, 
linearity and uniformity.

\begin{figure*}[t]
\centering
\includegraphics[width=8cm]{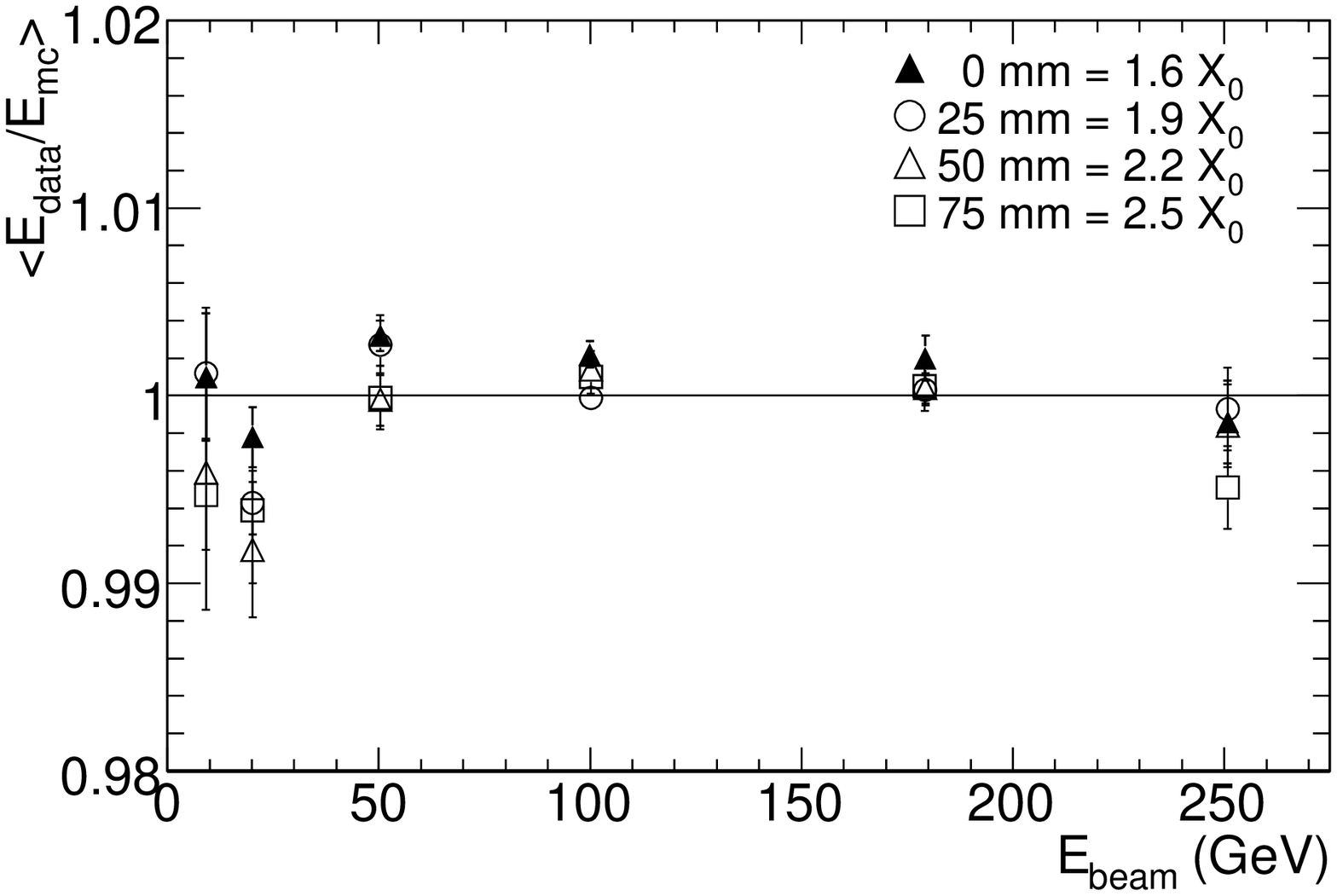}
\hspace{0.5cm}
\includegraphics[width=8cm]{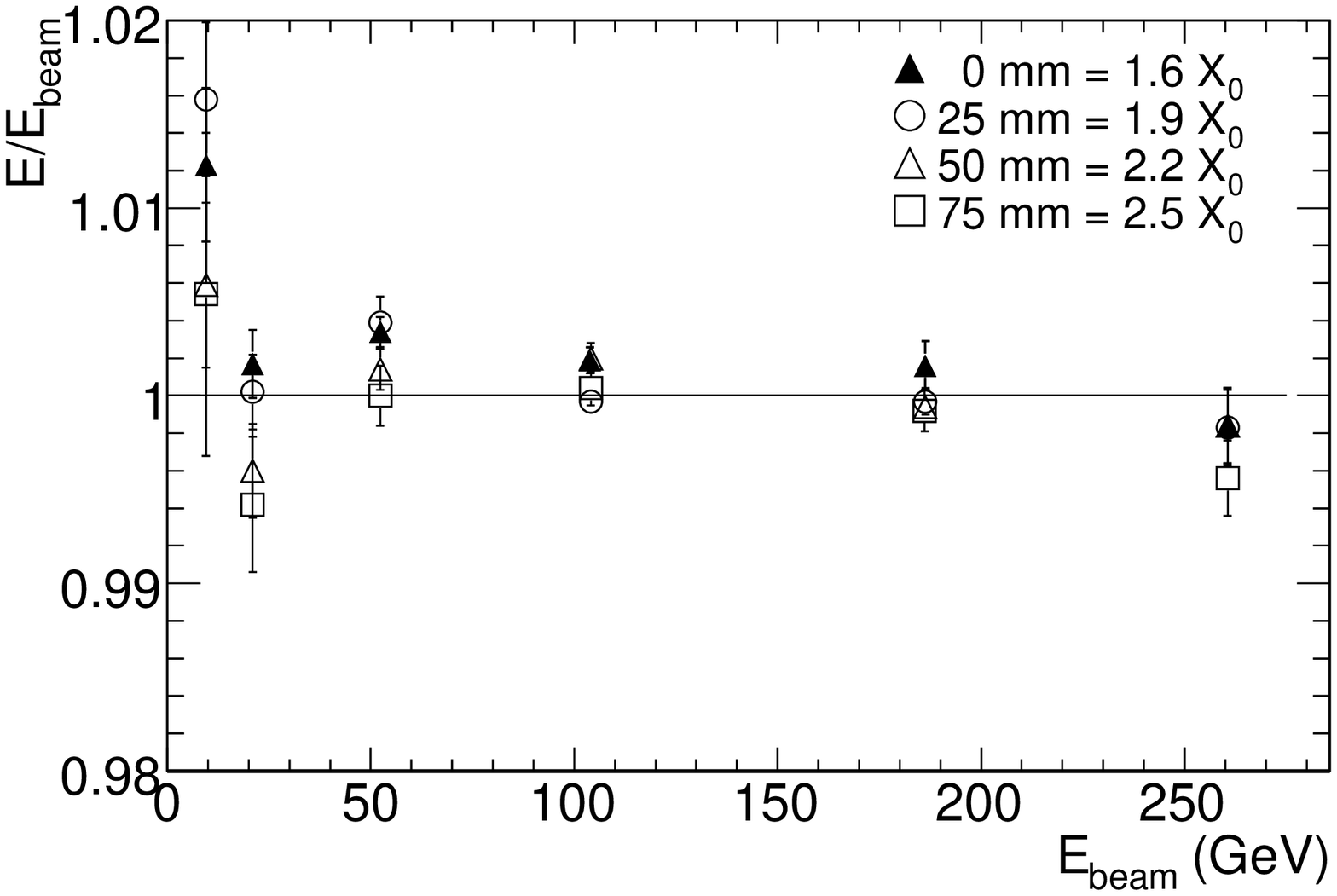}
\begin{picture}(0,0)
\put(-480, 160){a)}
\put(-220, 160){b)}
\end{picture}
\caption{a) Ratio of reconstructed over simulated electron
  cluster energy as function of beam energy, for 4 configurations of
  dead material. 
b) Ratio of reconstructed electron energy after calibration over beam
 energy for 4 configurations of dead material.} \label{fig-testbeam}
\end{figure*}

The ATLAS calibration strategy foresees to extract the calibration
parameters from a MC simulation. 
The energy calibration of the LAr calorimeter therefore relies
on precise MC simulations of the calorimeter response.
The reconstructed cluster energy for the
different layers of the EM calorimeter has been
studied for electrons from 1 to 250~GeV and for different amounts of
dead material in front of the calorimeter. Good agreement with the
simulation within 0.4\% was obtained (Fig.~\ref{fig-testbeam}a).
The level of agreement is limited by the systematic uncertainty of the beam energy measurement and 
the simulation of the beamline setup.

The energy linearity of the calorimeter response has been studied for
electron-beam energies between 1 and 250~GeV and for 4 different
configurations of dead material in front of the calorimeter. 
The observed non-linearity is of the order of 0.5\%
(Fig.~\ref{fig-testbeam}b). The main cause for the observed non-linearity 
is the description of the setup by the simulation. For a previous standalone
testbeam, where the simulation of the beamline setup was controlled
to very high precision, a non-linearity of 0.1\% was 
found~\cite{testbeam-linearity-resolution}, in agreement with the 
specification. 
The non-uniformity after applying
cell-level and cluster-level corrections was estimated for electron
beams up to 245~GeV to be approximately 0.5\%~\cite{testbeam-uniformity}. 

The energy resolution has been studied for different configurations of
dead material in front of the calorimeter. It was found that
the resolution worsens at a rate of approximately
0.5\%/$\sqrt{\mathrm{E}}$ per 30\% increase of a radiation length, $X_0$. Good agreement
between data and MC simulation was observed.

\section{CONCLUSIONS}
The installation of the LAr calorimeters has been completed and the
apparatus is now closed. 
Calibration measurements and cosmic muon data
are used successfully and continuously to commission and monitor the
LAr detector and its readout. Preliminary studies show 
only a small number of isolated permanently dead
readout channels of 0.02\%. All high voltage channels are operating,
less than 1\% of them at reduced voltage but with still usable
signals. 
Analysis from a combined test beam  
shows encouraging performance results. Good agreement between data and
MC simulation is observed. Linearity, uniformity and resolution are in
agreement with expectations.

\section{ACKNOWLEDGMENTS}
I would like to thank Manuella Vincter, Isabelle
Wingerter-Seez an Tancredi Carli for their support in preparing this document.


\begin{thebibliography}{9}   % Use for  1-9  references
%\begin{thebibliography}{99} % Use for 10-99 references

\bibitem{atlas-tdr}
ATLAS Collaboration,
% ``ATLAS: Detector and physics performance technical design report'',
%Volume 1, 
CERN-LHCC-99-14, 
%ATLAS-TDR-14, 
%Volume 2, 
CERN-LHCC-99-15, 
%ATLAS-TDR-15,
May 1999.

\bibitem{atlas-det}
ATLAS Collaboration, G. Aad et al., 
%``The ATLAS Experiment at the
%CERN Large Hadron Collider'',
 JINST 3 (2008) S08003. 

\bibitem{atlas-lar-tdr}
ATLAS Collaboration,
% ``ATLAS Liquid Argon Calorimeter Technical Design
%Report'', 
CERN/LHCC 96-41, ATLAS-TDR-2, December 1999.

\bibitem{atlas-calo-perf-tdr}
ATLAS Collaboration, 
%``ATLAS Calorimeter Performance Technical
%Design Report'', 
CERN/LHCC 96-40, ATLAS-TDR-1, January 1997
%, CERN,
%Geneva, Switzerland.

%\bibitem{fe-system}
%N.J.~Buchanan et al.,
%``ATLAS Liquid Argon Calorimeter Front End Electronics'', 
%accepted for publication in JINST.

\bibitem{feb-nim}
N.J.~Buchanan et al., 
%``Design and Implementation of the Front End
%Board for the Readout of the ATLAS Liquid Argon Calorimeters'',
2008~JINST~3~P03004, 2008.

\bibitem{backend}
J. Ban et al., 
%ATLAS liquid argon calorimeter back end electronics,
JINST 2 (2007) P06002. 

\bibitem{calib}
J. Colas et al.,
% Electronic calibration board for the ATLAS liquid argon calorimeter,
Nucl. Inst. Meth. A593 (2008) 269.

\bibitem{cosmic-note}
M.~Cook et al., 
%``In situ commissioning of the ATLAS electromagnetic
%calorimeter with cosmic muons'', 
ATL-LARG-PUB-2007-013, November 2007.

\bibitem{testbeam-linearity-resolution}
M.~Aharrouche et al., 
%``Energy Linearity and Resolution of the ATLAS
%Electromagnetic Barrel Calorimeter in an Electron Test-Beam'', 
Nucl. Instrum. Meth. A568 (2006) 601.

\bibitem{testbeam-uniformity}
M.~Aharrouche et al.,
% ``Response uniformity of the ATLAS liquid argon
%electromagnetic calorimeter'',
Nucl. Instrum. Meth. A582 (2007) 429.

\end{thebibliography}
\end{document}